\documentclass[onecolumn,referee,usenatbib]{mnras}

\usepackage{newtxtext,newtxmath}

\usepackage[T1]{fontenc}

\usepackage{graphicx}	
\usepackage{amsmath}	
\usepackage{chemfig}    
\usepackage[font=small,labelfont=bf]{caption}
\usepackage{lineno} 
\usepackage{siunitx} 
\usepackage[super]{nth}  

\title[Water Loss on Earth-like Exoplanets]{Higher Water Loss on Earth-like Exoplanets in Eccentric Orbits}

\author[Binghan Liu et al.]{
Binghan Liu,$^{1}$\thanks{E-mail: pybl@leeds.ac.uk}
Daniel R. Marsh,$^{1,2}$
Catherine Walsh$^{1}$
and Greg Cooke$^{1}$
\\
$^{1}$Department of Physics and Astronomy, University of Leeds, Woodhouse Ln., Leeds, LS2 9JT, UK \\
$^{2}$National Center for Atmospheric Research, Boulder CO 80301, USA
}

\date{Accepted XXX. Received YYY; in original form ZZZ}

\pubyear{2022}

\begin{document}
\label{firstpage}
\pagerange{\pageref{firstpage}--\pageref{lastpage}}
\maketitle

\begin{abstract}
The climate of a terrestrial exoplanet is controlled by the type of host star, the orbital configuration and the characteristics of the atmosphere and the surface. Many rocky exoplanets have higher eccentricities than those in the Solar System, and about 18\% of planets with masses $< 10~\mathrm{M}_{\oplus}$ have $e>0.1$. Underexplored are the implications of such high eccentricities on the atmosphere, climate, and potential habitability on such planets. We use WACCM6, a state-of-the-art fully-coupled Earth-system model, to simulate the climates of two Earth-like planets; one in a circular orbit ($e=0$), and one in an eccentric orbit ($e=0.4$) with the same mean insolation. We quantify the effects of eccentricity on the atmospheric water abundance and loss given the importance of liquid water for habitability. The asymmetric temperature response in the eccentric orbit results in a water vapour mixing ratio in the stratosphere ($> 20$~ppmv) that is approximately five times greater than that for circular orbit ($\sim 4$~ppmv). This leads to at most $\sim 3$ times increases in both the atmospheric hydrogen loss rate and the ocean loss rate compared with the circular case. Using the Planetary Spectrum Generator, we simulate the idealised transmission spectra for both cases. We find that the water absorption features are stronger at all wavelengths for the $e=0.4$ spectrum than for the circular case. Hence, highly-eccentric Earth-like exoplanets may be prime targets for future transmission spectroscopy observations to confirm, or otherwise, the presence of atmospheric water vapour.
\end{abstract}





\section{Introduction} 
\label{sec1}

An objective of the field of exoplanetary research is the hunt for extraterrestrial life. The potential for an exoplanet to harbour life is commonly assessed by its equilibrium temperature, assuming that the planet radiates as if it were a black body. A planet is deemed to be habitable if water is in a liquid state at this temperature; however, assessing the potential habitability of an exoplanet is less straightforward than the equilibrium temperature would suggest. The orbital parameters of a planet, such as eccentricity, play an important role in affecting the planetary climate. If we adopt 1.6 $\mathrm{R}_{\oplus}$ (Earth radii) and 10 $\mathrm{M}_{\oplus}$ (Earth masses) as the upper limits for defining rocky exoplanets, we find that among a total of 114 confirmed rocky exoplanets, 49 of them have measured eccentricities, and nine of them have an eccentricity greater than 0.1\footnote{The Extrasolar Planets Encyclopaedia: \url{http://exoplanet.eu/}}. For instance, TOI-1238 b has an eccentricity of 0.25, LTT 1445A c has an eccentricity of 0.223, and L 168-9 b has an eccentricity of 0.21 \citep{2022A&A...658A.138G,2022BAAS...54e.417W,2020A&A...636A..58A}. To date, GJ 1061 d, a potentially rocky exoplanet (> 1.5 $\mathrm{M}_{\oplus}$) detected by the radial velocity method, has the largest constrained eccentricity ($e<0.54$) \citep{2020MNRAS.493..536D}. The 10 $\mathrm{M}_{\oplus}$ sets the boundary between super-Earth and ice/gas giant planets, and beyond 1.6 $\mathrm{R}_{\oplus}$, the planet's density would be too low to be rocky \citep{2015ApJ...801...41R,2019A&A...631A..90L,2021arXiv211204663W}. Super-Earth and mini-Neptune exoplanets could vary largely in their planetary compositions, and the absence of their analogues in the Solar System makes it difficult to fully characterize them and determine if they are predominantly rocky. 

The climate response of rocky planets to changes in eccentricity has been investigated in previous work. \citet{2002IJAsB...1...61W} examined the climate of the current Earth with elliptical orbits near the habitable zone (HZ) of the Sun. They pointed out that long-term climate stability primarily depends on the average stellar flux received over an entire orbit, which is known as the mean flux approximation. However, later \citet{2016A&A...591A.106B} demonstrated that the mean flux approximation becomes less reliable for tidally-locked Earth-like planets in highly-eccentric orbits ($e \geq 0.6$) with host star luminosities between $10^{-4}$ $\mathrm{L_{\odot}}$ and 1 $\mathrm{L_{\odot}}$ as surface liquid water may not persist for the entire course of an orbit as the planet moves in and out of the habitable zone. The seasonality effects induced by higher eccentricities were considered by \citet{2010ApJ...721.1295D}, who found that Earth-like planets surrounding Sun-like stars do not necessarily suffer from long winters near aphelion due to thermal inertia if at least 10\% of the surface is ocean-covered. In addition, \citet{2015P&SS..105...43L} showed that for Earth-like planets, eccentric orbits extend the outer edge of the habitable zone and effectively limit the transition into snowball states. 

The time it takes for the planetary atmosphere to adapt to changes in radiation, also known as the atmospheric radiative timescale, can be affected by the mass of the atmosphere, the equilibrium temperature and the surface thermal inertia \citep{2014ClDy...43.1041D, 2019ApJ...881...67G,2023ApJ...943L...1J}. The seasonal response of temperature for rocky planets in eccentric orbits and with zero obliquity is strongly dependent on the interplay of the orbital period, rotation rate and radiative timescale \citep{2019AJ....157..189A,2020ApJ...901...46G,2022AGUA....300684G,2022ApJ...933...62H}. For example, the amplitude of the seasonal response decreases with decreasing rotation rate. For a constant rotation rate, longer orbital periods provide the atmosphere with more time to adjust to the changes in insolation, resulting in a stronger seasonal cycle. On the other hand, a longer radiative timescale means a weaker seasonal cycle because the atmosphere needs more time to respond to variations in insolation.

\citet{2019ApJ...874....1O} revealed that eccentricity might influence temperature patterns indirectly by affecting the radiative timescale; hence a transition from a diurnal mean insolation-controlled climate to an  annual mean insolation-controlled one is possible. Because equilibrium temperature increases with increasing eccentricity \citep{2022RNAAS...6...56Q}, we can deduce that, for a given atmospheric mass and orbital period, the orbital eccentricity can modulate the seasonal temperature response.


Eccentricity could also affect the habitability of rocky planets by affecting atmospheric loss rates. Originating near the planetary surface, hydrogen-bearing species in the form of \chemfig{H_{2}O}, \chemfig{H_{2}}, and \chemfig{CH_{4}} travel upward to the homopause through advection and turbulent mixing. In the upper homosphere, hydrogen-bearing molecules are dissociated by ultraviolet radiation, leaving hydrogen in atomic and molecular form only. \citet{1973JAtS...30.1481H} showed that the diffusion-limited hydrogen escape flux ($\Phi_{i}$), in units of molecules lost per unit time per unit area, is
\begin{flalign}\label{eqn1}
\Phi_{i}=\frac{b_i f_i}{H_a}
\end{flalign}
where $b_i$ is the binary diffusion parameter, $f_i$ is the mixing ratio of the hydrogen-bearing species at the homopause in units of moles per mole of air, and $H_a$ is the temperature-dependent atmospheric scale height at the homopause. The detailed derivation can be found in \cite{DavidCatling}. The atmospheric water concentration due to the effect of eccentricity has been examined by \cite{2017ApJ...835L...1W}, in which they identify temperate climatic conditions when varying the eccentricity from 0 to 0.283. They found that the tropopause water vapour mixing ratio is the highest at $e=0.283$, but it is still nearly 15 times lower than the moist-greenhouse limit\footnote{The moist-greenhouse limit occurs when water vapour starts accumulating in the stratosphere with a mixing ratio $ > 10^{-3}$.} at perihelion. In addition, \citet{2020ApJ...890...30P} showed ocean worlds with higher eccentricity orbits are more likely to lose water, and that if the eccentricity is greater than 0.55 for an Earth-like aqua-planet orbiting a G-type star, the whole water inventory will be lost owing to the runaway greenhouse effect. 


The effects of eccentricity on water loss owing to an increased solar constant have been studied using both 1D radiative-convective models (RCMs) \citep[e.g.][]{1984Icar...57..335K,2015ApJ...813L...3K} and 3D general circulation models (GCMs) \citep[e.g.][]{2015JGRD..120.5775W,2015P&SS..105...43L,2017ApJ...845....5K,2020ApJ...901...46G}. They find a general trend that water loss increases with increasing solar constant. While there are a few papers which addressed the climate seasonality effect due to increasing eccentricity with the mean flux approximation for Earth-like exoplanets \citep[e.g.][]{2002IJAsB...1...61W,2016A&A...591A.106B}, they did not quantify the atmospheric water vapour abundance with varying eccentricity. In addition, the 3D GCMs used in \cite{2002IJAsB...1...61W} and \cite{2016A&A...591A.106B} are not coupled with whole atmosphere chemistry, and the atmosphere's vertical extent does not reach the homopause at which the diffusion-limited escape of hydrogen can be estimated. In this study, we quantify the water abundance and estimate the water loss rate for a highly-eccentric rocky exoplanet using the fully-coupled whole-atmosphere Earth-system model, WACCM 6 (Section \ref{sec2.1}). We describe the simulation configurations in Section \ref{sec2.2}. The simulation results are shown in Section \ref{sec3}. The impact of our results on the potential observability using the Planetary Spectrum Generator (PSG) is included in the Discussion (Section \ref{sec4}). In Section \ref{sec5}, we summarize our findings.

\section{Methods} \label{sec2}

\subsection{Model Description} \label{sec2.1}

Simulations are performed using WACCM6 (the Whole Atmosphere Community Climate Model version 6), which is a configuration in CESM2.2 (the Community Earth System Model version 2.2) \citep{2019JGRD..12412380G}. WACCM6, an updated version from WACCM4 \citep{2013JCli...26.7372M}, is a high-top 3D atmosphere model used primarily for studying the pre-industrial, present day and potential future climates of Earth. As an interactive Earth-system model, WACCM6 consists of a land model with soil, vegetation and topography, an ocean model with ocean dynamics and sea-ice, and an atmosphere model with fully coupled chemistry. WACCM6 is capable of studying the Earth's climate with whole atmosphere chemistry and dynamics from the surface ($\sim 1000$ hPa) to the lower thermosphere ($6\times 10^{-6}$~hPa) with two horizontal resolution options of $1^{\circ}$ and $2^{\circ}$.

The chemical mechanism which we opt for our simulations is the middle atmosphere chemistry scheme (MA) with a horizontal resolution of $2^{\circ}$, which is a subset of the comprehensive troposphere-stratosphere-mesosphere-lower thermosphere chemistry scheme (TSMLT) \citep{2019JGRD..12412380G,2020JAMES..1201882E}. MA includes 97 chemical species, 208 chemical reactions and 90 photolysis reactions. Compared with TSMLT, MA requires much less computational resources owning to a reduced set of tropospheric reactions (i.e., prescribed sulfate aerosols and neglected non-methane hydrocarbons). The radiative transfer code is the Rapid Radiative Transfer Model for GCMs (RRTMG) using the correlated-K approach \citep{1997JGR...10216663M,2008JGRD..11313103I,2012GMD.....5..709L,2019JGRD..12412380G} in which the line integration over discrete wave number is replaced by the integration over correlated continuous cumulative probability density function. The parameterizations of boundary layer, shallow convection and cloud macrophysics are performed using the Cloud Layers Unified By Binormals (CLUBB) in which the small scale variabilities are predicted by the multivariate probability density function \citep{2002JAtS...59.3519L,2013JCli...26.9655B,2022JAMES..1403127L}. Cloud microphysics parameterizations uses the Morrison-Gettelman Version 2 microphysics (MG2) which explicitly predicts the mixing ratios and number concentrations of cloud water droplet and cloud ice \citep{2014JCli...27.6821P,2015JCli...28.1268G}. The aerosol treatment is the Modal Aerosol Model Version 4 (MAM4) in which aerosol size distributions are represented by multiple lognormal functions \citep{2012GMD.....5..709L,2016GMD.....9..505L,2016JGRD..121.2332M}. WACCM6 has been used extensively to study the Earth's climate from the ancient geologic era to pre-industrial and modern times \citep[e.g.][]{2022JGRD..12736452Z,2022WtFor..37..797R,2022GeoRL..4999848D}. Recently, the model was used to study the habitability and climate of Earth-like exoplanets. For example, \cite{2022MNRAS.tmp.2450C,2022RSOS....911165C} calculated \chemfig{O_3} column variations at different \chemfig{O_2} concentrations based on knowledge of the ancient Earth, and the implications of variable \chemfig{O_2} concentrations on observations of oxygenated Earth-analogue exoplanets.

\subsection{Simulation Configurations} \label{sec2.2}

We initialize our simulations with equilibrated conditions on January \nth{1} in the year 1850 (so-called pre-industrial era or PI) with a mean surface temperature of $287.2~\mathrm{K}$. The PI controlled case has cyclic lower boundary conditions for the zonally symmetric surface chemical emissions, which annually repeats the same amplitude and the same seasonal variation for each chemical species. For instance, the mixing ratios of the greenhouse gases on the surface, such as $\chemfig{CO_2}$ $\chemfig{CH_4}$ and $\chemfig{N_2O}$, have an annual global mean of $284$ ppmv, $0.808$ ppmv and $0.273$ ppmv, respectively. The seasonal variations of these greenhouse gases are $\sim 1\%$ . Other H species, such as $\chemfig{H_2}$, have surface emissions fixed at $0.5$ ppmv in latitude throughout the year. The major atmospheric constitutes in the PI controlled case are about $21\%$ $\chemfig{O_2}$ and $78\%$ $\chemfig{N_2}$.

After running the PI case for 101 days when the total solar irradiance (TSI) equals its annual mean, we branch into  two different simulations with eccentricities of $e=0$ and $e=0.4$. The orbital obliquity is set to be zero, and the longitude of perihelion is fixed at $100.29^{\circ}$ in both cases. To focus on the climate response due to varying eccentricity only, we fix the annual mean insolation of the $e=0.4$ case (the highly-eccentric case) and the $e=0$ (the circular case) to the same value as for the Earth's PI case ($1360.75~\mathrm{W~m}^{-2}$). Using Eq.~\ref{eqn2} and Eq.~\ref{eqn3}, the TSI is $1248.58~\mathrm{W~m}^{-2}$ for the $e=0.4$ case, and $1360.94~\mathrm{W~m}^{-2}$ for the circular case using the flux ratio calculated according to \cite{1978JAtS...35.2362B}. 
\begin{flalign}
TSI_{e=0.4}&=TSI_{PI} \cdot \frac{F_{e=0.4}}{F_{e=0.0167}}\label{eqn2}\\
TSI_{e=0}  &=TSI_{PI} \cdot \frac{F_{e=0}}{F_{e=0.0167}}\label{eqn3}
\end{flalign}

We verify that a quasi-steady state has been reached by running the model until the trends in the Earth Energy Imbalance (EEI) for the $e=0.4$ and $e=0$ cases between two consecutive years are well within their inter-annual variability. It took eight simulation years for the $e=0.4$ case to reach a quasi-steady state, whereas the $e=0$ case reached a quasi-steady state after only one simulation year due to a smaller change in the eccentricity from the PI case. All simulation data is averaged over five simulation years after the quasi-steady state is reached to average over inter-annual variability and to avoid the influence of the model spin-up time due to the change in eccentricity. 


Two important modifications were made in the simulation configurations compared with a default PI run. Firstly, we mute the QBO (Quasi-biennial Oscillation) forcing because it is a specified tuning parameter which was originally included to reproduce the Earth's stratospheric water vapour distribution. Secondly, we disable MARBL\footnote{MARBL is a sub-model that controls marine ecosystem dynamics and the coupled cycles of carbon, nitrogen, phosphorus, iron, silicon, and oxygen \citep{2021JAMES..1302647L}.} (Marine Geo-biochemistry Library) in the model configurations to simplify the calculations and because it is also tailored to Earth's ecosystem, which we cannot assume is the same on other planets.

\section{RESULTS}  \label{sec3}

\subsection{Temperature and Water Vapor Profiles} \label{sec3.1}

\begin{figure}
    \centering
    \includegraphics[width=0.8\columnwidth]{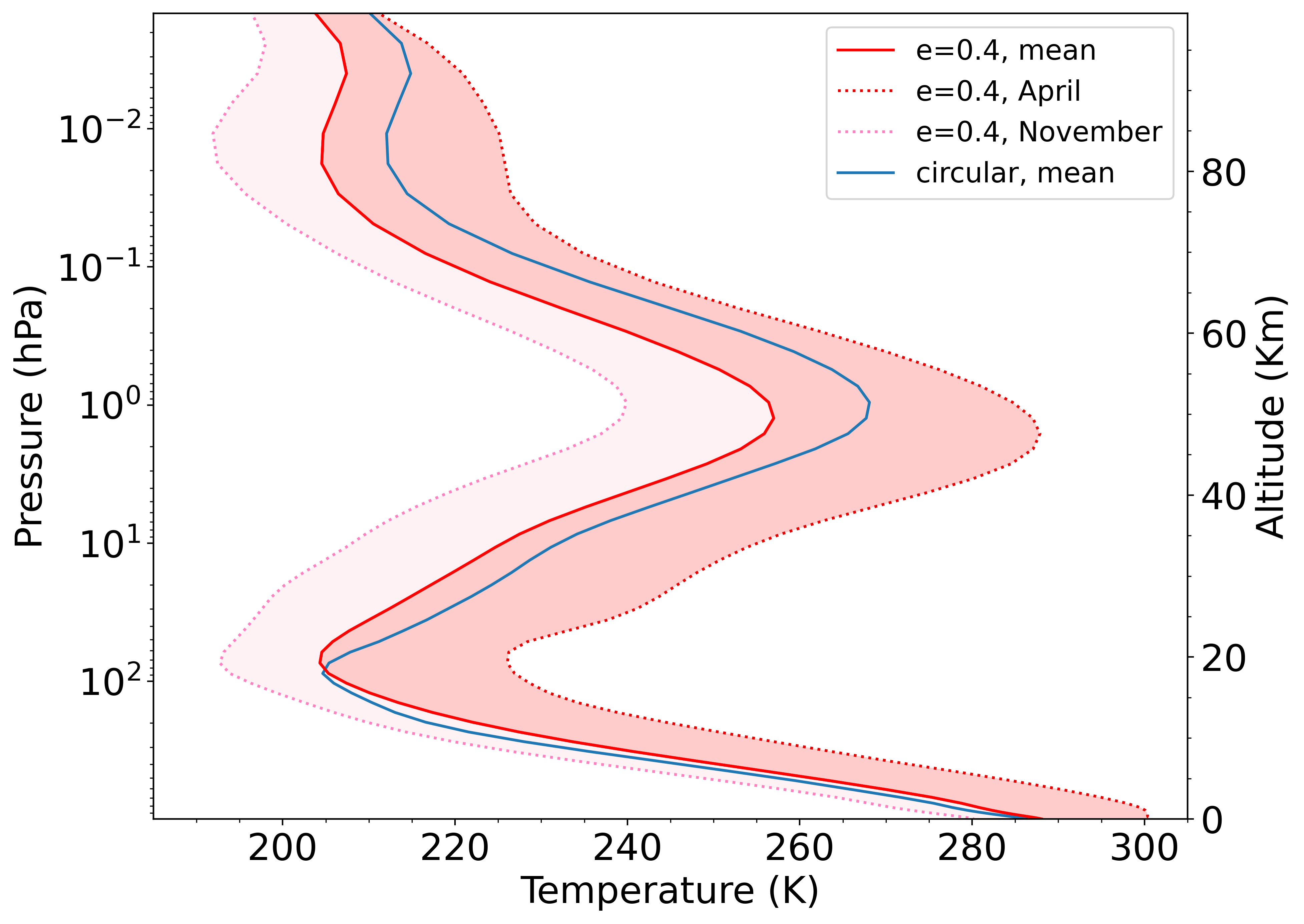}
    \caption{\large The annual mean atmospheric temperature profile for the $e=0.4$ case (red) and the circular case (blue) from the surface to the homopause. The hot (red shaded area) and cold (pink shaded area) passages for the $e=0.4$ case cover the time when the atmospheric temperature is higher than the mean and is lower than the mean, respectively. The temperature profiles for the hottest month (April) and the coldest month (November) are indicated by the dotted lines on the outer boundary of the shaded areas. For demonstration purposes, we give a secondary y-axis to indicate the altitude in km for the circular case. For the $e=0.4$ case, the corresponding altitude in km does not deviate much from the secondary y-axis because they have similar atmospheric scale heights ($\delta H \sim 0.5$~km).}
    \label{fig1}
\end{figure}

Figure \ref{fig1} shows the globally-averaged annual mean vertical temperature profiles for the circular case (blue) and the $e=0.4$ case (red). The temperature decreases with increasing altitude in the troposphere up to the tropopause, at which a temperature inversion occurs. We show the globally-averaged monthly mean temperature profiles to demonstrate the range in temperature experienced by the eccentric planet in April (red dotted line) and November (pink dotted line) at which the highest and lowest surface temperatures are reached, respectively, and to compare with the annual means. The annual mean temperature profile for the $e=0.4$ case has a slightly warmer troposphere with a $1.3$~K higher surface temperature than the circular case. The tropopause is moved up in altitude in the $e=0.4$ case but its tropopause temperature is lower than the circular case by 1.2~K. In the upper atmosphere from the tropopause to the mesopause, the annual mean temperature in the $e=0.4$ case is lower by 6.7 K on average than in the circular case, and the maximum temperature difference of 11.1 K occurs at the stratopause. That  the stratosphere is colder may be due to a 3 $\%$ less annual mean ozone column due to increased production of OH radicals from increased water vapor photolysis (see Section \ref{sec4} for more details). Less ozone means fewer UV absorbers, and hence they contribute less heating to the stratosphere. In addition, the increase in water vapor in the stratosphere could have a radiative cooling effect \citep{2001GeoRL..28.2791O}.

Although the stratosphere and mesosphere are, on average, colder in the eccentric case, there is significant seasonal variation in the temperature profiles which reaches a maximum in April and a minimum in November due to the lagged response of surface and tropospheric temperature. It takes roughly two months for both the planetary surface and the atmosphere to respond to the highest insolation at perihelion, and three months to respond to the minimum insolation at aphelion (see Fig. \ref{fig1}). The lag may correspond to the radiative timescale of the coupling between the surface and the atmosphere \citep{2013JAMES...5..843C}. A simplified approach from \citet{2020ApJ...901...46G} states that the radiative timescale, $\tau$, can be approximated as $$\tau \propto C Q^{-3/4}$$ where $C$ is the atmospheric heat capacity and $Q$ is the incoming insolation at the top of the atmosphere. Assuming that $C$ remains approximately constant over the eccentric orbit, the radiative timescale should decrease with increasing $Q$. This is consistent with what we see here which is that the surface and tropospheric temperature lagged response at aphelion is longer than at perihelion. Furthermore, the surface temperature difference between the peak and the annual mean ($|\delta TS|=12.8$ K) is significantly greater than the difference between the trough and the annual mean ($|\delta TS|=7.8$~K). Thus, the lagged response of surface and tropospheric temperature for the $e=0.4$ case is seasonally asymmetric in both the response time and the response magnitude. 

\begin{figure}
    \centering
    \includegraphics[width=0.8\columnwidth]{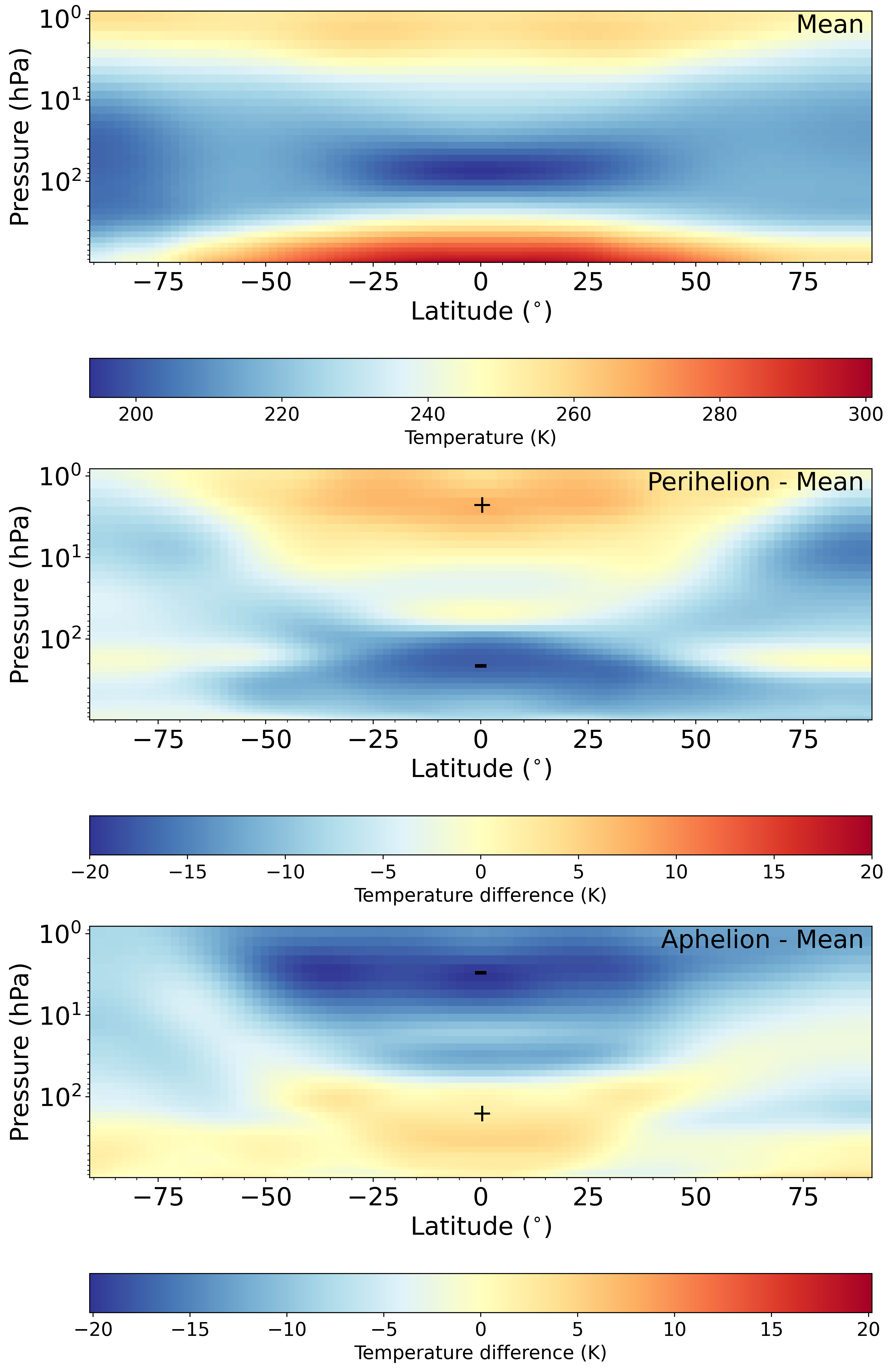}
    \caption{\large Upper panel: annual, zonal mean of air temperature from the surface to the stratopause over the last five simulation years after steady state is reached for the $e=0.4$ case. Middle panel: the difference in air temperature between perihelion and the mean for the $e=0.4$ case. Bottom panel: the difference in air temperature between aphelion and the mean for the $e=0.4$ case. Note that the top panel has a different colour scale than the middle and bottom panels. The plus and minus signs in the figure denote the positive and the negative temperature differences, respectively.}
    \label{fig2}
\end{figure}

\begin{figure}
    \centering
    \includegraphics[width=0.8\columnwidth]{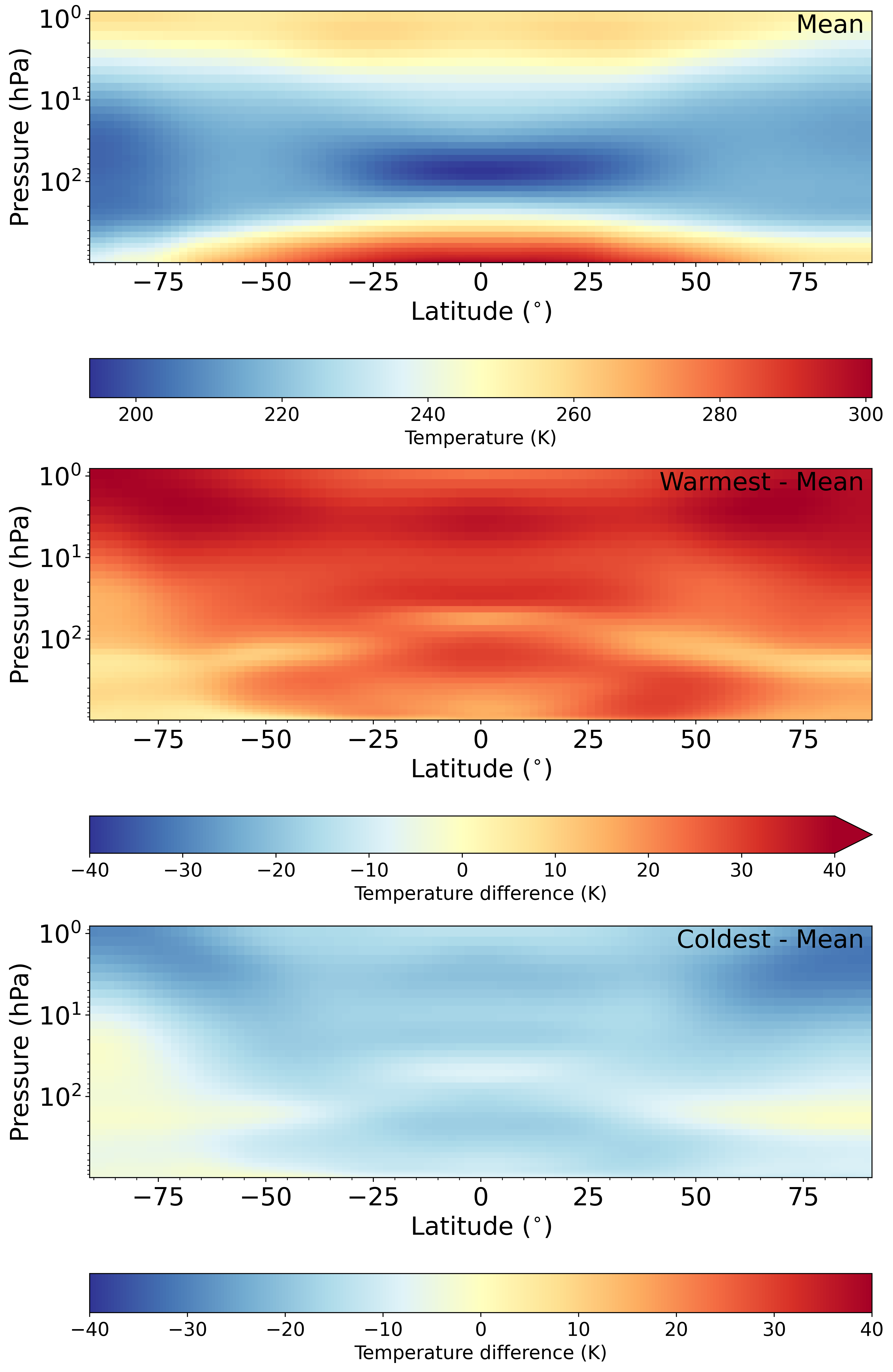}
    \caption{\large Upper panel: annual, zonal mean of air temperature from the surface to the stratopause over the last five simulation years after steady state is reached for the $e=0.4$ case. Middle panel: the difference in air temperature between the warmest month (April) and the mean for the $e=0.4$ case. Bottom panel: the difference in air temperature between the coldest month (November) and the mean for the $e=0.4$ case. Note that the top panel has a different colour scale than the middle and bottom panels.}
    \label{fig3}
\end{figure}

The top panel of Fig. \ref{fig2} shows the annual mean temperature (top panel) from the surface to the stratopause as a function of latitude for the $e=0.4$ case, and the bottom two panels show the temperature difference at perihelion (middle panel) and aphelion (bottom panel) relative to the annual mean temperature profile. The mean temperature pattern reflects the fact that Earth's obliquity is set to zero such that the zonal mean surface temperature decreases between the tropics and the poles. Similar to Earth, the Arctic region is warmer at the surface than the Antarctic region because the former is ocean-covered and hence has a larger heat capacity mitigating the cold winters in the eccentric orbit. Unlike the whole atmospheric and surface warming and cooling during the hottest month (April) and the coldest month (November) shown in the middle and bottom panel of Fig.~\ref{fig3}, respectively, one feature in Fig.~\ref{fig2} is the atmospheric and surface lagged response to the change in insolation, as shown in the bottom panel. At perihelion where the planet receives 2.64 times its mean insolation, the troposphere ($<10^{2}$~hPa) is colder than the annual mean temperature. However, the temperature pattern is reversed in the stratosphere between $10^{2}$ to 1 hPa. We find that the stratospheric temperature is lagged by 1 month relative to perihelion, and 2 months relative to aphelion, whereas the tropospheric temperature is better coupled with the surface temperature. In addition to the ocean heat buffer at aphelion, that the troposphere and the surface are warmer in the $e=0.4$ case can be attributed to a weaker cloud shortwave radiative effect, which contributes about 5 $\mathrm{Wm^{-2}}$ more into the system compared to the circular case. The cloud shortwave effect controls the transmitted radiation from the cloud layer down to the surface. The atmospheric and surface lagged response to insolation helps the climate to remain temperate in these two extreme scenarios.

\begin{figure}
    \centering
    \includegraphics[width=0.8\columnwidth]{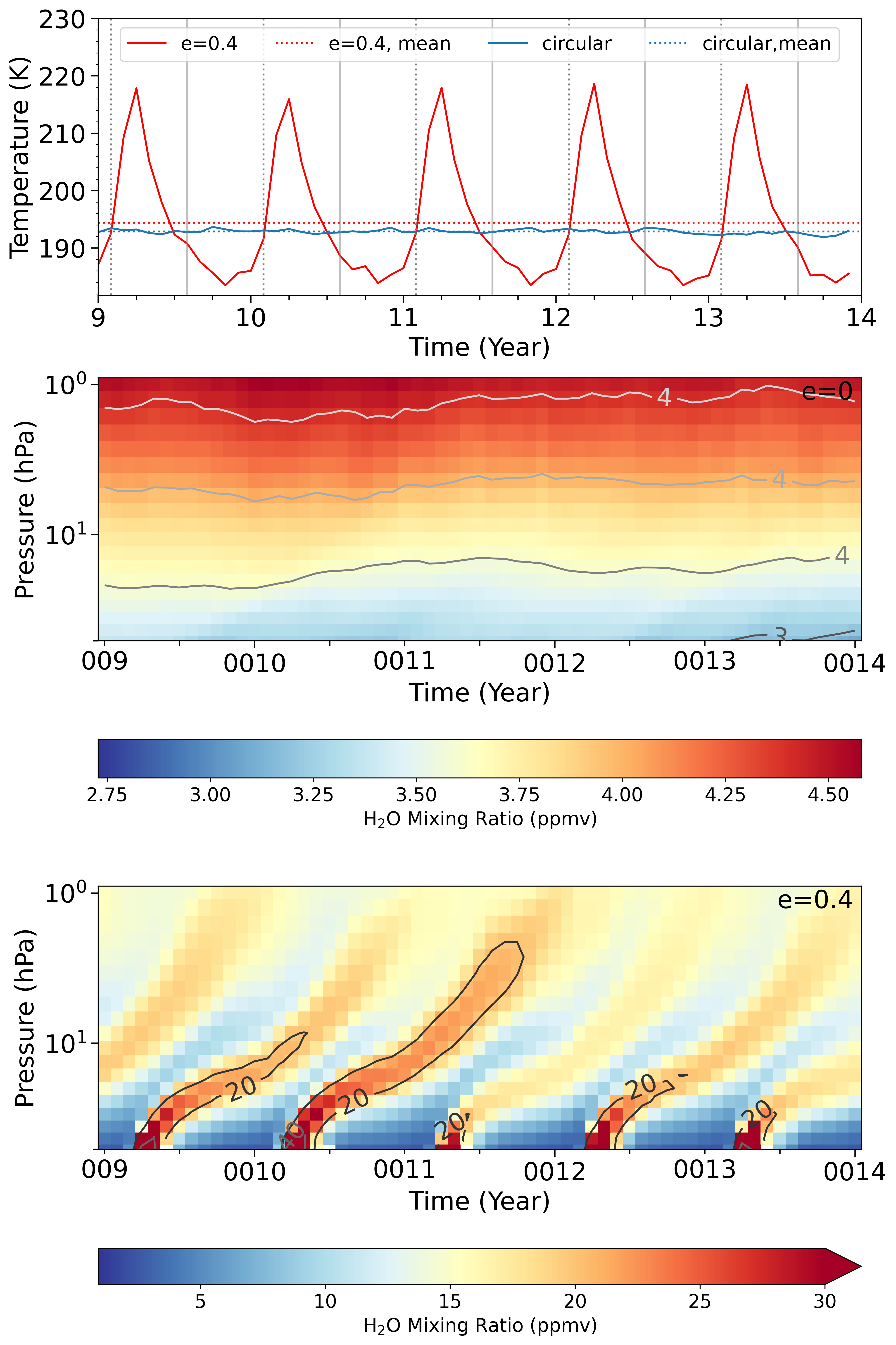}
    \caption{\large Upper panel: the seasonal variation in the tropical tropopause temperatures for the $e=0.4$ case (red solid line) and the circular case (blue solid line). Also shown are the annual mean of the tropical tropopause temperatures for the $e=0.4$ case (red dotted line) and the circular case (blue dotted line) over the last five consecutive years of the simulations. Perihelion takes place in February (grey dotted line), and aphelion takes place six months afterwards in August (grey solid line). Middle and lower panels: the \protect\chemfig{H_{2}O} mixing ratio in the stratosphere for the circular case and the $e=0.4$ case over the last five consecutive years of the simulation, respectively. Note the presence of the `tape recorder' signal in the bottom plot for the $e=0.4$ case. The black contours highlight a constant mixing ratio of 20~ppmv and the `tape recorder' signal is moving vertically upward in each case. Note that the middle and bottom panels have different colour scales, and the dark red regions in the bottom panels are over-saturated with a water vapour mixing ratio from 30 ppmv to 80 ppmv as indicated by the arrow on the right-hand of the colour bar.}
    \label{fig4}
\end{figure}

During the hot orbital passage, more surface water can be evaporated and uplifted into the tropical troposphere as the surface temperature rises. Through condensation, the tropical tropopause, also known as the cold trap, confines most of the rising water vapour in the form of clouds, but a small amount of water vapour can escape into the stratosphere \footnote{Moist air could bypass the cold trap for high obliquity worlds \citep{2019ApJ...877L...6K}}. The stratospheric water vapour concentration oscillates seasonally at the same frequency as the tropical tropopause temperature, which is known as the `tape recorder' signal. Here, we examine the `tape recorder' signal in the circular and $e=0.4$ cases. The upper panel in Fig. \ref{fig4} shows the seasonal variation of tropical tropopause temperature for the $e=0.4$ case (red) and the circular case (blue). The corresponding stratospheric water vapour mixing ratios in ppmv for these two cases are shown in the middle and bottom panels of Fig. \ref{fig4}. The tropical tropopause altitude is calculated for each month over the last five simulation years after steady state is reached since the tropopause can be lifted upward or downward depending on the surface temperature in the $e=0.4$ case. As indicated by the dotted lines in the top panel of Fig. \ref{fig4}, the annual mean tropical tropopause temperatures are $194.27$~K in the $e=0.4$ case and $193.07$~K in the circular case. Although the annual mean temperatures differ by $\sim 1$~K between these two cases, the seasonal variation of the tropical tropopause temperature in the $e=0.4$ case is significantly higher than in the circular case. From the middle and the lower panels, the water vapour mixing ratios at the bottom of the stratosphere for both cases follow the same temporal pattern as shown in their tropical tropopause temperature profiles. This also indicates that the water vapour mixing ratios at the cold trap are in phase with the surface and tropospheric temperature. In April, the temperature is over $\sim 215$~K in the $e=0.4$ case, in contrast to $\sim 193$~K in the circular case. The cold trap in April for the $e=0.4$ case becomes less effective, allowing a larger amount of water vapour (up to 30 ppmv) to enter the stratosphere, producing a clear tape recorder signal.

\section{DISCUSSION} \label{sec4}

\subsection{Water Loss Rate and Ocean Loss Timescale}

\begin{figure}
    \centering
    \includegraphics[width=0.8\columnwidth]{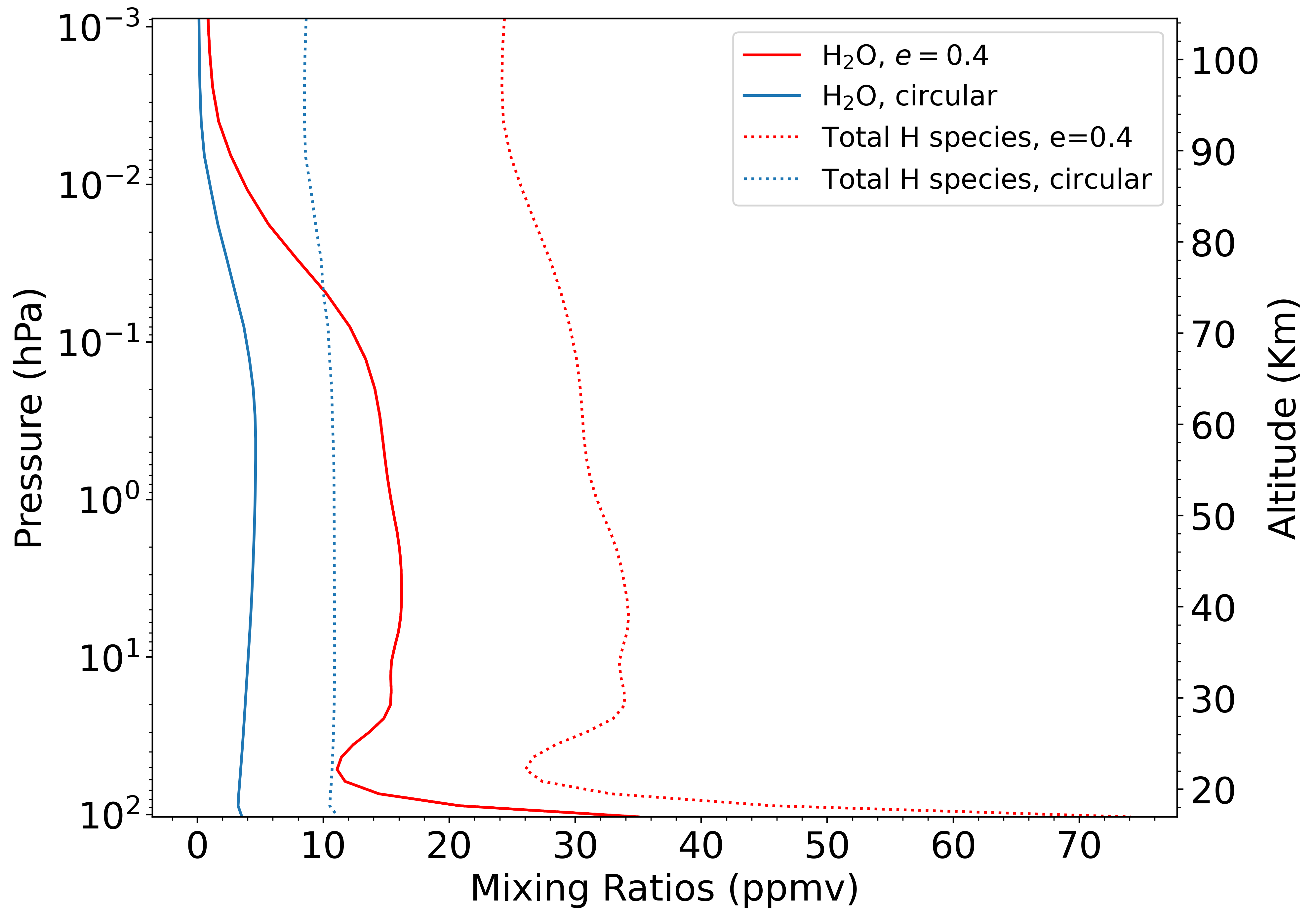}
    \caption{\large The global, annual mean \protect \chemfig{H_{2}O} and total \protect\chemfig{H} species mixing ratios in the stratosphere averaged over the last 5 simulation years for the $e=0.4$ case (red solid and red dotted line, respectively) and the circular case (blue solid and blue dotted lines, respectively). The vertical scale in the plot is set to cover from the tropopause to the homopause for the circular case. For demonstration purposes, we give a secondary y-axis which is used to indicate the altitude in km for the circular case only. For the $e=0.4$ case, the corresponding altitude in km does not deviate much from the secondary y-axis because both cases have similar atmospheric scale heights ($\delta H \sim$ 0.5 km).}
    \label{fig5}
\end{figure}

Figure \ref{fig5} shows the \chemfig{H_{2}O} and total \chemfig{H} species ($\chemfig{H}+2\cdot\chemfig{H_{2}}+2\cdot\chemfig{H_{2}O}+4\cdot\chemfig{CH_{4}}$) mixing ratios. For simplification, we assume that the total hydrogen species comprises only the four listed species, as these are the primary hydrogen carriers in the upper atmosphere of modern Earth. The annual mean mixing ratio of the tropical tropopause water vapour for the $e=0.4$ case (red dotted line) is 11.3~ppmv and is about three times greater than that for the circular case (blue dotted line), though it is still about two orders of magnitude lower than the threshold of $3 \times 10^3$~ppmv needed to trigger moist greenhouse states \citep[e.g.][]{1984Icar...57..335K, 1988Icar...74..472K}. 

To quantify how quickly water can escape into space, we compute the total loss rate at the homopause, in units of atoms~cm$^{-2}$~s$^{-1}$, as the sum of individual hydrogen-bearing species as follows: 
\begin{flalign} \label{eqn4}
\Phi_H =\frac{1}{H_a}\cdot (b_H f_{H} + 2\cdot b_{H_2}f_{H_2}+2\cdot b_{H_{2}O} f_{H_{2}O} + 4\cdot{b_{CH_4}f_{CH_4}})
\end{flalign} where $f$ is the mixing ratio of the selected hydrogen carrier in ppmv, $b$ is the binary diffusion parameter in units of \SI{}{cm^{-1}~s^{-1}}, and $H_{a}$ is the atmospheric scale height in cm. The scale height is $H_{a} = \frac{k_{B}T_{hp}}{mg}$ where $k_B$ is Boltzmann's constant, $T_{hp}$ is the globally averaged homopause temperature, $m$ is the mean atmospheric molecular mass, and $g$ is the surface gravity. The binary diffusion parameter formulas can be found in \cite{1973JAtS...30.1481H} for $b_\mathrm{H}$, $b_{\mathrm{H}_{2}}$, $b_{\mathrm{H}_{2}\mathrm{O}}$, and in \cite{Banks&Kockarts} for $b_{\mathrm{CH}_{4}}$, and we reproduce them here:
\begin{flalign} 
b_{H}&=6.5\times 10^{17} \cdot T_{hp}^{0.7} \nonumber\\
b_{\mathrm{H}_{2}}&= 2.67 \times 10^{17} \cdot T_{hp}^{0.75} \nonumber\\ b_{\mathrm{H}_{2}\mathrm{O}}&= 0.137\times 10^{17} \cdot T_{hp}^{1.072} \nonumber\\ b_{\mathrm{CH}_{4}}&= 0.756\times 10^{17} \cdot T_{hp}^{0.747} \nonumber\\
\end{flalign} where $T_{hp}$ is the homopause temperature in K.

The homopause is taken to be at a pressure level of $1.458 \times 10^{-3}$~hPa ($\sim 100$~km), and the total hydrogen species mixing ratios at homopause in the $e=0.4$ case (red dotted line in Fig. \ref{fig5}) is $24.2$~ppmv.  This is about three times larger than the $8.5$~ppmv calculated for the circular case (blue dotted line). The escape fluxes calculated using the above equations are $6.33\times10^8$ and $2.31 \times 10^8$ $\mathrm{atoms~cm^{-2}~s^{-1}}$ for the $e=0.4$ case and the circular case, respectively. Additionally, we have compared the escape fluxes from taking the global annual mean operator over the individual component on the right hand side of Eq.~\ref{eqn4} against taking the global annual mean operator over the sum of all species, and we find a $0.3\%$ difference in the escape flux. Thus, it can be confirmed that this uncertainty is not significant. To compare with previous estimates, the H escape flux in the circular case is lower than that \citet{DavidCatling} calculated for the Earth's eccentricity case ($3.5 \times 10^8$ $\mathrm{atoms~cm^{-2}~s^{-1}}$). This could be due to the obliquity and eccentricity differences, which result in a lower homopause temperature, a smaller scale height, and a lower total hydrogen species mixing ratio when calculating the escape flux. In addition, the hydrogen mixing ratio values in that work are taken from the lower stratosphere instead of the homopause. Thus, any bottlenecks that could exist above the lower stratosphere may not have been included in their calculations (see Fig. \ref{fig5} for the differences in the total H mixing ratio at different altitudes). 

We can use our calculated \chemfig{H} escape fluxes to estimate the ocean survival timescale for both cases. The total mass of water in Earth's ocean is $1.4 \times 10^{24}$ \SI{}{g} \citep[e.g.][]{2015JGRD..120.5775W,2017ApJ...845....5K}. We compute the corresponding total hydrogen reservoir to be about $9.36 \times 10^{46}$ $\mathrm{atoms}$ so that the ocean loss timescale is 2,437 Gyr for the circular case and
891 Gyr for the $e=0.4$ case. As a result, an Earth-like planet in an eccentric orbit ($e=0.4$) around a Sun-like star loses its water inventory in approximately one third of the time it takes a planet in a circular orbit to lose its water inventory. Despite this higher ocean loss rate, our results suggest that a highly-eccentric Earth-like exoplanet could retain its oceans over the lifetime of the stellar system. Our result agrees with \citet{2020ApJ...890...30P} who predict, using a 1D model, that an Earth-like exoplanet around a G-type star in an $e=0.4$ orbit has a dry upper atmosphere (i.e., one that is far away from entering a runaway greenhouse state). However, it is worth noting that complete ocean water loss has been predicted to occur within 2 Gyr for our Earth under the brightening Sun due to its evolution through the main sequence stage \citep{2015JGRD..120.5775W}. Thus, the ocean loss timescale under a brightening host star should be much shorter than what we have estimated here for both the circular and the eccentric cases. How both orbital eccentricity and evolving host star luminosity together affect the water loss should be tested in future simulations. On the other hand, our water loss rate estimates are upper bounds under constant insolation as they are computed based on the mixing ratio of the total \chemfig{H} species which can enter into the homopause \citep{1974JAtS...31..305H}. So, the ocean loss timescales calculated in the two cases here could be longer depending on the efficiencies of active escape mechanisms (i.e., thermal escape, non-thermal escape, impact erosion). We note also that the water loss rate on an Earth-like exoplanet is also likely to be influenced by the stellar type of the host star \citep[e.g.][]{2017ApJ...837..107W,2021ApJ...909L...2K}. Moreover, the ocean loss timescale also depends on the initial water inventory, and so it is possible for an Earth-like exoplanet with a much smaller initial water reservoir ($<$\SI{1}{\%}) to lose all of its water content within the lifetime of its stellar system.



\subsection{Implications for Observations}

\subsubsection{Calculating the spectra}

Using the Planetary Spectrum Generator (PSG;\cite{2018JQSRT.217...86V}) with an idealized telescope and a constant spectral resolving power of 250, we calculated the transit spectrum in the UV, visible and infrared regions between 0.2 to \SI{20}{\micro \metre} to determine if the water abundances predicted for both of our cases leads to potentially observable differences in the synthetic transit spectra. In PSG, the layer-by-layer radiative transfer is done by the Planetary and Universal Model of Atmospheric Scattering (PUMAS; \cite{2018JQSRT.217...86V}), and the line-by-line intensity calculation uses molecular cross sections from the latest HITRAN database \citep{2022JQSRT.27707949G}. The geometry of the observation is set such that the planet is in front of the host star at a phase of $180^{\circ}$ for all cases. So, the planet's terminator is located at $90^{\circ}$ and $270^{\circ}$ in longitude. The inclination angle is fixed at $90^{\circ}$ (edge-on) for all cases. For the eccentric case, we take snapshots of the climate on April \nth{16} and November \nth{12} to represent the hottest and coldest days, respectively. The planet-star separation of the $e=0.4$ case is 1.03~au on April \nth{16}, and 1.2~au on November \nth{12}. For the circular case, the planet-star separation is fixed at 1~au, and we take one snapshot on April \nth{16} only for comparison since the climate experiences no seasonal variation (recall that both simulations assume zero obliquity). Note that we simulate spectra only at two distinct points in the orbit representative of two extreme cases (corresponding to the hottest and coldest day). Whilst we may expect simulated spectra from other orbital positions to lie within these two scenarios, this should be tested with further simulations.

The atmospheric composition is assumed to be made of \chemfig{N_2}, \chemfig{H_2O}, \chemfig{O_2}, \chemfig{O_3}, \chemfig{CO_2}, \chemfig{N_2O}, \chemfig{CH_4} as they are the dominant absorbers expected over our chosen wavelength range. We feed PSG with the instantaneous WACCM6 output which includes the pressure-temperature structure, the mixing ratios of the molecules in the atmosphere, cloud fraction, and cloud ice fraction. Model data were re-binned to a resolution of $5.625^{\circ}$ in latitude (i.e., a binning number equal to 3) to save costs for computing the radiative transfer. Hence, a total of 32 sample points in latitude on the terminator are averaged over instead of the original 96 latitudinal points from the WACCM6 runs.


\subsubsection{Synthetic Transit Spectra}

\begin{figure}
    \centering
    \includegraphics[width=0.8\columnwidth]{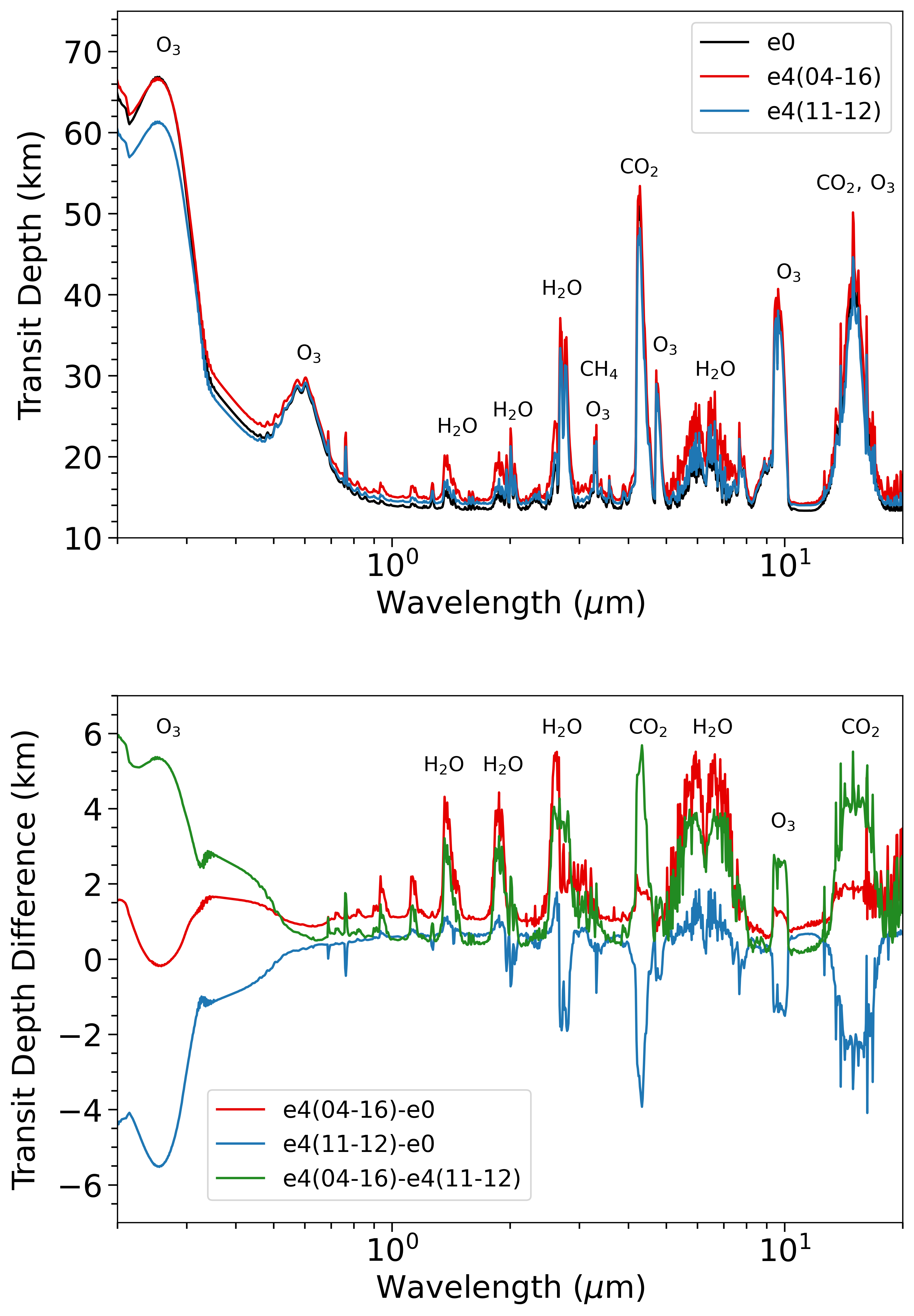}
    \caption{\large Upper panel: the transit spectra of the circular case (black), the snapshot on April \nth{16} of the $e=0.4$ case (red) and the snapshot on November \nth{12} of the $e=0.4$ case (blue) between \SI{0.2}{\mu m} to \SI{20}{\mu m}; Bottom panel: the transit depth difference between the snapshot on April \nth{16} of the $e=0.4$ case and the circular case (red), the transit depth difference between the snapshot on November \nth{12} of the $e=0.4$ case and the circular case (blue) and the transit depth difference between the snapshots on April \nth{16} and November \nth{12} of the $e=0.4$ case (green).}
    \label{fig6}
\end{figure}

The upper panel of Fig. \ref{fig6} shows the synthetic transit spectra for the circular case (black), the snapshot on April \nth{16} for the $e=0.4$ case (red) and the snapshot on November \nth{12} of the $e=0.4$ case (blue). The lower panel shows the differences in transit depth between the circular case and the snapshot on April \nth{16} of the $e=0.4$ case (red) and between the circular case and the snapshot on November \nth{12} of the $e=0.4$ case (blue), and between the snapshots on April \nth{16} and November \nth{12} of the $e=0.4$ case (green). Absorption signatures from $\chemfig{H_2O}$, $\chemfig{O_3}$, $\chemfig{CO_2}$ and $\chemfig{CH_4}$ are present in the synthetic spectra. The strong absorption feature near \SI{4.8}{\micro \metre} is due to $\chemfig{CO_2}$ with a transit depth of \SI{55}{km}. For water, the dominant feature is at \SI{2.6}{\micro \metre} with a $\sim 37$~\SI{}{km} transit depth. The largest $\chemfig{O_3}$ feature with a transit depth of \SI{65}{km} is at around \SI{0.25}{\micro \metre}. The feature at \SI{15}{\micro \metre} has a transit depth of \SI{50}{km} and is due to both \chemfig{CO_2} and \chemfig{O_3}. The baselines of the three cases are different due to a combined effect of differences in the planet-star separation, cloud coverage and temperature. In general, a larger planet-star separation, more cloud coverage and higher temperature all increase the baseline \citep{2014ApJ...791....7B,2019ApJ...887..194F}. The effect of clouds is important in the $e=0.4$ case because there is more cloud than in the circular case, and the cloud extends vertically to the lower stratosphere below \SI{20}~{hPa}. Thus, the cloud in the $e=0.4$ case can, not only increase the transit depth of the baseline, but also decrease the depth of spectral features in the infrared.

In the lower panel of Fig. \ref{fig6}, the water features on April \nth{16} are about 5 km deeper than on November \nth{12} (green line), reflecting the largest seasonal difference in the water abundance in the $e=0.4$ case. This is seen in the water absorption features present at \SI{1.36}{\micro \metre}, \SI{1.87}{\micro \metre}, \SI{2.7}{\micro \metre} and between \SI{5.93}{\micro \metre} and \SI{6.64}{\micro \metre}. The difference in the depth of the water features between the April \nth{16} snapshot in the $e=0.4$ case and the circular case is 1 to 2 km larger (red line) than the maximum seasonal difference. The differences in the depth of the water features between the November \nth{12} snapshot in the $e=0.4$ case and the circular case (blue line) is the smallest of our three comparisons; however, slightly stronger water absorption features are present even on the coldest day of the $e=0.4$ case. The reason for this is related to the strong release of water vapor into the stratosphere during the warmest month (April in the $e=0.4$ case) which slowly propagates upward in the stratosphere (see the bottom panel Fig. \ref{fig4}). Thus, the water vapor column above the baseline at a given time depends on both the current and the previous release of water vapor above the cold trap. This is essentially a result of the strong seasonality induced by the large eccentricity. 

The stronger water features on April \nth{16} compared to November \nth{12} and the circular case can be explained by the higher temperature in the April \nth{16} snapshot for the $e=0.4$ case. A higher surface temperature increases the evaporation rate of \chemfig{H_2O}, which increases the \chemfig{H_2O} number density with altitude; more \chemfig{H_2O} molecules mean stronger absorption in the infrared. Some secondary effects could also contribute to the transit depth difference. For example, the \chemfig{H_2O} absorption cross-section positively correlates to the temperature at these wavelengths \citep{2022JQSRT.27707949G}. The higher temperature on April \nth{16} also causes a more inflated atmosphere, and hence more water vapour molecules are present at a higher altitude. 

Unlike the \chemfig{H_2O} features which are always stronger in the highly-eccentric case than the circular case, the \chemfig{CO_2} transit depth features near \SI{4.8}{\micro \metre} and \SI{15}{\micro \metre} are 2 to 3 km deeper for the April \nth{16} case than the circular case, but they are about 4 km deeper for the circular case than on November \nth{12}. The seasonal variation of the \chemfig{CO_2} transit features is about 6 km between the April and the November cases. The reasons for these are the same as just described for the water. The atmosphere expands at higher temperatures, and the \chemfig{CO_2} density remains high at higher altitudes than if the atmosphere were cooler.

The ozone layer in the stratosphere is an important atmospheric constituent on Earth as it can prevent harmful UV radiation from reaching the surface. The ozone concentration in the atmosphere does not vary as much as the water vapour in the highly-eccentric system. The annual mean ozone column density decreases by about 3\% in the $e=0.4$ case compared to the circular case, and the seasonal ozone column density in the $e=0.4$ case decreases up to about 13\% in April compared to the circular case. This is affected by the water vapour abundance in the atmosphere: a higher water vapour abundance leads to more \chemfig{OH} radicals formed in the atmosphere via photolysis, and hence more ozone is destroyed by the increasing amount of \chemfig{OH} radicals. The seasonal variation of \chemfig{O_3} density in the stratospheric ozone layer for the $e=0.4$ case ($\sim~16$\%) could impact the temporal surface habitability \citep{2019Icar..321..608G}. In the upper panel of Fig. \ref{fig6}, the most prominent ozone features of the transit spectrum peaks around \SI{0.25}{\micro \metre} in the UV region, \SI{0.6}{\micro \metre} in the visible, and around \SI{9.6}{\micro \metre} in the infrared. There are small \chemfig{O_3} absorption depth differences between the April \nth{16} snapshot case and the circular case, whilst there are noticeable \chemfig{O_3} absorption depth differences ($\sim$ \SI{2}{km}) between the April \nth{16} snapshot case and the November \nth{12} snapshot case. 

Contrary to expectations, the $e=0.4$ case's ozone features are stronger on April \nth{16} than on November \nth{12}, as shown in the lower panel of Fig. \ref{fig6}, despite the former having a lower ozone column density than the latter. Similarly, when comparing the circular case and the November \nth{12} snapshot case, we find that a higher ozone column density does not correspond to a more pronounced ozone feature. This is because a higher atmospheric temperature causes larger ozone absorption cross-sections \citep{amt-7-625-2014,2022JQSRT.27707949G}, and the \chemfig{O_3} number density above 40~km is higher in both the April \nth{16} snapshot case and the circular case than in the November \nth{12} snapshot case. Therefore, future transmission spectra observations could result in degenerate interpretations for inferring the total ozone column density for Earth-like exoplanets.


\section{CONCLUSIONS} \label{sec5}

Terrestrial exoplanets can have large orbital eccentricities up to values as high as $0.4$ to $0.5$ \citep{2019A&A...631A..90L}. The climate response of rocky planets to changes in eccentricity has been studied in recent years using both 1D models and 3D general circulation models (GCMs). Previous work has examined water loss due to an increasing solar constant from both Earth and other potential exoplanet host stars. However, whether the eccentricity-induced seasonality effect alone has a critical impact on water loss had yet to be studied. Using the fully coupled 3D Earth climate model, WACCM6, we simulated the climate response of Earth-like rocky planets around a Sun-like star that have different orbital eccentricities. We compared the temperature structures and the water loss rates calculated for a circular orbit case and a highly-eccentric case ($e=0.4$). We find that the water loss rate for a planet in the $e=0.4$ orbit is about 3 times larger than if it were in a circular orbit under the same annual mean insolation. Consequently, the ocean loss timescale for an Earth-like exoplanet in such an eccentric orbit is about 3 times shorter than the circular orbit. Moreover, an Earth-like planet with an eccentricity of $0.4$ spends half of its orbit outside of the Earth's habitable zone as traditionally defined by \cite{1993Icar..101..108K}. Nevertheless, we find that the climate remains temperate and the planet can hold on to its water reservoir. In addition, Earth-like exoplanets with a high orbital eccentricity could have stronger \chemfig{H_2O} and \chemfig{CO_2} absorption features in simulated transmission spectroscopy than for the same planet in a circular orbit. Hence, the climate response to eccentricity as indicated by a higher water abundance may be measurable with future observations. Stronger \chemfig{O_3} absorption features seen in future observations may not always imply a higher ozone column density for an Earth-like exoplanet with an ozone layer in its atmosphere. It is worth noting the limitations of this study. First of all, the model we adopted in the simulation uses numerous parameterizations that are tuned to Earth’s climate and might be inappropriate to generalize to an exoplanetary climate with different conditions. Hence, these conclusions should be tested in future simulations with more flexible chemistry, land and ocean coverage, and cloud and radiative transfer schemes. Secondly, a variety of effects such as QBO and ocean bio-geochemistry are excluded from the simulation, though they will not change our results qualitatively. In addition, the Earth-like exoplanet's atmosphere evolution over its geological history and the effects of space weather have been ignored because these factors are beyond the scope of this study. For a more comprehensive view of the habitability analysis of Earth-like exoplanets, future work should include a systemic climate study due to the change of eccentricity with more effects included and more inter-comparisons between different 1D and 3D model simulations.

\clearpage
\newpage
\section*{Acknowledgements}

This work was undertaken on ARC4, part of the High Performance Computing facilities at the University of Leeds, UK. C.W.~acknowledges financial support from the University of Leeds, the Science and Technology Facilities Council, and UK Research and Innovation (grant numbers ST/T000287/1 and MR/T040726/1). G.J.C.~acknowledges the studentship funded by the Science and Technology Facilities Council of the United Kingdom (STFC; grant number ST/T506230/1). 
\section*{Data Availability}

The data underlying this article are now available in the Dryad Digital Repository, at \url{doi:10.5061/dryad.jsxksn0f5}.

\clearpage
\newpage



\bibliographystyle{mnras}
\bibliography{example} 

\begin{thebibliography}{}
\makeatletter
\relax
\def\mn@urlcharsother{\let\do\@makeother \do\$\do\&\do\#\do\^\do\_\do\%\do\~}
\def\mn@doi{\begingroup\mn@urlcharsother \@ifnextchar [ {\mn@doi@}
  {\mn@doi@[]}}
\def\mn@doi@[#1]#2{\def\@tempa{#1}\ifx\@tempa\@empty \href
  {http://dx.doi.org/#2} {doi:#2}\else \href {http://dx.doi.org/#2} {#1}\fi
  \endgroup}
\def\mn@eprint#1#2{\mn@eprint@#1:#2::\@nil}
\def\mn@eprint@arXiv#1{\href {http://arxiv.org/abs/#1} {{\tt arXiv:#1}}}
\def\mn@eprint@dblp#1{\href {http://dblp.uni-trier.de/rec/bibtex/#1.xml}
  {dblp:#1}}
\def\mn@eprint@#1:#2:#3:#4\@nil{\def\@tempa {#1}\def\@tempb {#2}\def\@tempc
  {#3}\ifx \@tempc \@empty \let \@tempc \@tempb \let \@tempb \@tempa \fi \ifx
  \@tempb \@empty \def\@tempb {arXiv}\fi \@ifundefined
  {mn@eprint@\@tempb}{\@tempb:\@tempc}{\expandafter \expandafter \csname
  mn@eprint@\@tempb\endcsname \expandafter{\@tempc}}}

\bibitem[\protect\citeauthoryear{{Adams}, {Boos}  \& {Wolf}}{{Adams}
  et~al.}{2019}]{2019AJ....157..189A}
{Adams} A.~D.,  {Boos} W.~R.,   {Wolf} E.~T.,  2019, \mn@doi [\aj]
  {10.3847/1538-3881/ab107f}, \href
  {https://ui.adsabs.harvard.edu/abs/2019AJ....157..189A} {157, 189}

\bibitem[\protect\citeauthoryear{{Astudillo-Defru} et~al.,}{{Astudillo-Defru}
  et~al.}{2020}]{2020A&A...636A..58A}
{Astudillo-Defru} N.,  et~al., 2020, \mn@doi [\aap]
  {10.1051/0004-6361/201937179}, \href
  {https://ui.adsabs.harvard.edu/abs/2020A&A...636A..58A} {636, A58}

\bibitem[\protect\citeauthoryear{Banks \& Kockarts}{Banks \&
  Kockarts}{1973}]{Banks&Kockarts}
Banks Kockarts 1973, AERONOMY (PART B).
ACADEMIC PRESS, LIBRARY OF CONGRESS CATALOG CARD NUMBER: 72-88332

\bibitem[\protect\citeauthoryear{{Berger}}{{Berger}}{1978}]{1978JAtS...35.2362B}
{Berger} A.~L.,  1978, \mn@doi [Journal of Atmospheric Sciences]
  {10.1175/1520-0469(1978)035<2362:LTVODI>2.0.CO;2}, \href
  {https://ui.adsabs.harvard.edu/abs/1978JAtS...35.2362B} {35, 2362}

\bibitem[\protect\citeauthoryear{{B{\'e}tr{\'e}mieux} \&
  {Kaltenegger}}{{B{\'e}tr{\'e}mieux} \&
  {Kaltenegger}}{2014}]{2014ApJ...791....7B}
{B{\'e}tr{\'e}mieux} Y.,  {Kaltenegger} L.,  2014, \mn@doi [\apj]
  {10.1088/0004-637X/791/1/7}, \href
  {https://ui.adsabs.harvard.edu/abs/2014ApJ...791....7B} {791, 7}

\bibitem[\protect\citeauthoryear{{Bogenschutz}, {Gettelman}, {Morrison},
  {Larson}, {Craig}  \& {Schanen}}{{Bogenschutz}
  et~al.}{2013}]{2013JCli...26.9655B}
{Bogenschutz} P.~A.,  {Gettelman} A.,  {Morrison} H.,  {Larson} V.~E.,  {Craig}
  C.,   {Schanen} D.~P.,  2013, \mn@doi [Journal of Climate]
  {10.1175/JCLI-D-13-00075.1}, \href
  {https://ui.adsabs.harvard.edu/abs/2013JCli...26.9655B} {26, 9655}

\bibitem[\protect\citeauthoryear{{Bolmont}, {Libert}, {Leconte}  \&
  {Selsis}}{{Bolmont} et~al.}{2016}]{2016A&A...591A.106B}
{Bolmont} E.,  {Libert} A.-S.,  {Leconte} J.,   {Selsis} F.,  2016, \mn@doi
  [\aap] {10.1051/0004-6361/201628073}, \href
  {https://ui.adsabs.harvard.edu/abs/2016A&A...591A.106B} {591, A106}

\bibitem[\protect\citeauthoryear{Catling \& Kasting}{Catling \&
  Kasting}{2017}]{DavidCatling}
Catling Kasting 2017, Atmospheric Evolution on Inhabited and Lifeless Worlds.
Cambridge University Press, ISBN 9781139020558

\bibitem[\protect\citeauthoryear{{Cooke}, {Marsh}, {Walsh}, {Rugheimer}  \&
  {Villanueva}}{{Cooke} et~al.}{2022a}]{2022MNRAS.tmp.2450C}
{Cooke} G.~J.,  {Marsh} D.~R.,  {Walsh} C.,  {Rugheimer} S.,   {Villanueva}
  G.~L.,  2022a, \mn@doi [\mnras] {10.1093/mnras/stac2604}, \href
  {https://ui.adsabs.harvard.edu/abs/2022MNRAS.tmp.2450C} {}

\bibitem[\protect\citeauthoryear{{Cooke}, {Marsh}, {Walsh}, {Black}  \&
  {Lamarque}}{{Cooke} et~al.}{2022b}]{2022RSOS....911165C}
{Cooke} G.~J.,  {Marsh} D.~R.,  {Walsh} C.,  {Black} B.,   {Lamarque} J.~F.,
  2022b, \mn@doi [Royal Society Open Science] {10.1098/rsos.211165}, \href
  {https://ui.adsabs.harvard.edu/abs/2022RSOS....911165C} {9, 211165}

\bibitem[\protect\citeauthoryear{{Cronin} \& {Emanuel}}{{Cronin} \&
  {Emanuel}}{2013}]{2013JAMES...5..843C}
{Cronin} T.~W.,  {Emanuel} K.~A.,  2013, \mn@doi [Journal of Advances in
  Modeling Earth Systems] {10.1002/jame.20049}, \href
  {https://ui.adsabs.harvard.edu/abs/2013JAMES...5..843C} {5, 843}

\bibitem[\protect\citeauthoryear{{Donohoe}, {Frierson}  \&
  {Battisti}}{{Donohoe} et~al.}{2014}]{2014ClDy...43.1041D}
{Donohoe} A.,  {Frierson} D. M.~W.,   {Battisti} D.~S.,  2014, \mn@doi [Climate
  Dynamics] {10.1007/s00382-013-1843-4}, \href
  {https://ui.adsabs.harvard.edu/abs/2014ClDy...43.1041D} {43, 1041}

\bibitem[\protect\citeauthoryear{{Dreizler} et~al.,}{{Dreizler}
  et~al.}{2020}]{2020MNRAS.493..536D}
{Dreizler} S.,  et~al., 2020, \mn@doi [\mnras] {10.1093/mnras/staa248}, \href
  {https://ui.adsabs.harvard.edu/abs/2020MNRAS.493..536D} {493, 536}

\bibitem[\protect\citeauthoryear{{Dressing}, {Spiegel}, {Scharf}, {Menou}  \&
  {Raymond}}{{Dressing} et~al.}{2010}]{2010ApJ...721.1295D}
{Dressing} C.~D.,  {Spiegel} D.~S.,  {Scharf} C.~A.,  {Menou} K.,   {Raymond}
  S.~N.,  2010, \mn@doi [\apj] {10.1088/0004-637X/721/2/1295}, \href
  {https://ui.adsabs.harvard.edu/abs/2010ApJ...721.1295D} {721, 1295}

\bibitem[\protect\citeauthoryear{{Dub{\'e}}, {Randel}, {Bourassa}  \&
  {Degenstein}}{{Dub{\'e}} et~al.}{2022}]{2022GeoRL..4999848D}
{Dub{\'e}} K.,  {Randel} W.,  {Bourassa} A.,   {Degenstein} D.,  2022, \mn@doi
  [\grl] {10.1029/2022GL099848}, \href
  {https://ui.adsabs.harvard.edu/abs/2022GeoRL..4999848D} {49, e99848}

\bibitem[\protect\citeauthoryear{{Emmons} et~al.,}{{Emmons}
  et~al.}{2020}]{2020JAMES..1201882E}
{Emmons} L.~K.,  et~al., 2020, \mn@doi [Journal of Advances in Modeling Earth
  Systems] {10.1029/2019MS001882}, \href
  {https://ui.adsabs.harvard.edu/abs/2020JAMES..1201882E} {12, e2019MS001882}

\bibitem[\protect\citeauthoryear{{Fauchez} et~al.,}{{Fauchez}
  et~al.}{2019}]{2019ApJ...887..194F}
{Fauchez} T.~J.,  et~al., 2019, \mn@doi [\apj] {10.3847/1538-4357/ab5862},
  \href {https://ui.adsabs.harvard.edu/abs/2019ApJ...887..194F} {887, 194}

\bibitem[\protect\citeauthoryear{{Gettelman} \& {Morrison}}{{Gettelman} \&
  {Morrison}}{2015}]{2015JCli...28.1268G}
{Gettelman} A.,  {Morrison} H.,  2015, \mn@doi [Journal of Climate]
  {10.1175/JCLI-D-14-00102.1}, \href
  {https://ui.adsabs.harvard.edu/abs/2015JCli...28.1268G} {28, 1268}

\bibitem[\protect\citeauthoryear{{Gettelman} et~al.,}{{Gettelman}
  et~al.}{2019}]{2019JGRD..12412380G}
{Gettelman} A.,  et~al., 2019, \mn@doi [Journal of Geophysical Research
  (Atmospheres)] {10.1029/2019JD030943}, \href
  {https://ui.adsabs.harvard.edu/abs/2019JGRD..12412380G} {124, 12,380}

\bibitem[\protect\citeauthoryear{{G{\'o}mez-Leal}, {Kaltenegger}, {Lucarini}
  \& {Lunkeit}}{{G{\'o}mez-Leal} et~al.}{2019}]{2019Icar..321..608G}
{G{\'o}mez-Leal} I.,  {Kaltenegger} L.,  {Lucarini} V.,   {Lunkeit} F.,  2019,
  \mn@doi [\icarus] {10.1016/j.icarus.2018.11.019}, \href
  {https://ui.adsabs.harvard.edu/abs/2019Icar..321..608G} {321, 608}

\bibitem[\protect\citeauthoryear{{Gonz{\'a}lez-{\'A}lvarez}
  et~al.,}{{Gonz{\'a}lez-{\'A}lvarez} et~al.}{2022}]{2022A&A...658A.138G}
{Gonz{\'a}lez-{\'A}lvarez} E.,  et~al., 2022, \mn@doi [\aap]
  {10.1051/0004-6361/202142128}, \href
  {https://ui.adsabs.harvard.edu/abs/2022A&A...658A.138G} {658, A138}

\bibitem[\protect\citeauthoryear{{Gordon} et~al.,}{{Gordon}
  et~al.}{2022}]{2022JQSRT.27707949G}
{Gordon} I.~E.,  et~al., 2022, \mn@doi [\jqsrt] {10.1016/j.jqsrt.2021.107949},
  \href {https://ui.adsabs.harvard.edu/abs/2022JQSRT.27707949G} {277, 107949}

\bibitem[\protect\citeauthoryear{{Guendelman} \& {Kaspi}}{{Guendelman} \&
  {Kaspi}}{2019}]{2019ApJ...881...67G}
{Guendelman} I.,  {Kaspi} Y.,  2019, \mn@doi [\apj] {10.3847/1538-4357/ab2a06},
  \href {https://ui.adsabs.harvard.edu/abs/2019ApJ...881...67G} {881, 67}

\bibitem[\protect\citeauthoryear{{Guendelman} \& {Kaspi}}{{Guendelman} \&
  {Kaspi}}{2020}]{2020ApJ...901...46G}
{Guendelman} I.,  {Kaspi} Y.,  2020, \mn@doi [\apj] {10.3847/1538-4357/abaef8},
  \href {https://ui.adsabs.harvard.edu/abs/2020ApJ...901...46G} {901, 46}

\bibitem[\protect\citeauthoryear{{Guendelman} \& {Kaspi}}{{Guendelman} \&
  {Kaspi}}{2022}]{2022AGUA....300684G}
{Guendelman} I.,  {Kaspi} Y.,  2022, \mn@doi [AGU Advances]
  {10.1029/2022AV000684}, \href
  {https://ui.adsabs.harvard.edu/abs/2022AGUA....300684G} {3, e2022AV000684}

\bibitem[\protect\citeauthoryear{{He}, {Merrelli}, {L'Ecuyer}  \&
  {Turnbull}}{{He} et~al.}{2022}]{2022ApJ...933...62H}
{He} F.,  {Merrelli} A.,  {L'Ecuyer} T.~S.,   {Turnbull} M.~C.,  2022, \mn@doi
  [\apj] {10.3847/1538-4357/ac6951}, \href
  {https://ui.adsabs.harvard.edu/abs/2022ApJ...933...62H} {933, 62}

\bibitem[\protect\citeauthoryear{{Hunten}}{{Hunten}}{1973}]{1973JAtS...30.1481H}
{Hunten} D.~M.,  1973, \mn@doi [Journal of Atmospheric Sciences]
  {10.1175/1520-0469(1973)030\textless{}1481:TEOLGF\textgreater{}2.0.CO;2},
  \href {https://ui.adsabs.harvard.edu/abs/1973JAtS...30.1481H} {30, 1481}

\bibitem[\protect\citeauthoryear{{Hunten} \& {Strobel}}{{Hunten} \&
  {Strobel}}{1974}]{1974JAtS...31..305H}
{Hunten} D.~M.,  {Strobel} D.~F.,  1974, \mn@doi [Journal of Atmospheric
  Sciences] {10.1175/1520-0469(1974)031<0305:PAEOTH>2.0.CO;2}, \href
  {https://ui.adsabs.harvard.edu/abs/1974JAtS...31..305H} {31, 305}

\bibitem[\protect\citeauthoryear{{Iacono}, {Delamere}, {Mlawer}, {Shephard},
  {Clough}  \& {Collins}}{{Iacono} et~al.}{2008}]{2008JGRD..11313103I}
{Iacono} M.~J.,  {Delamere} J.~S.,  {Mlawer} E.~J.,  {Shephard} M.~W.,
  {Clough} S.~A.,   {Collins} W.~D.,  2008, \mn@doi [Journal of Geophysical
  Research (Atmospheres)] {10.1029/2008JD009944}, \href
  {https://ui.adsabs.harvard.edu/abs/2008JGRD..11313103I} {113, D13103}

\bibitem[\protect\citeauthoryear{{Ji}, {Bailey}, {Fabrycky}, {Kite}, {Jiang}
  \& {Abbot}}{{Ji} et~al.}{2023}]{2023ApJ...943L...1J}
{Ji} X.,  {Bailey} N.,  {Fabrycky} D.,  {Kite} E.~S.,  {Jiang} J.~H.,   {Abbot}
  D.~S.,  2023, \mn@doi [\apjl] {10.3847/2041-8213/acaf62}, \href
  {https://ui.adsabs.harvard.edu/abs/2023ApJ...943L...1J} {943, L1}

\bibitem[\protect\citeauthoryear{{Kaltenegger} \& {Lin}}{{Kaltenegger} \&
  {Lin}}{2021}]{2021ApJ...909L...2K}
{Kaltenegger} L.,  {Lin} Z.,  2021, \mn@doi [\apjl] {10.3847/2041-8213/abe634},
  \href {https://ui.adsabs.harvard.edu/abs/2021ApJ...909L...2K} {909, L2}

\bibitem[\protect\citeauthoryear{{Kang}}{{Kang}}{2019}]{2019ApJ...877L...6K}
{Kang} W.,  2019, \mn@doi [\apjl] {10.3847/2041-8213/ab1f79}, \href
  {https://ui.adsabs.harvard.edu/abs/2019ApJ...877L...6K} {877, L6}

\bibitem[\protect\citeauthoryear{{Kasting}}{{Kasting}}{1988}]{1988Icar...74..472K}
{Kasting} J.~F.,  1988, \mn@doi [\icarus] {10.1016/0019-1035(88)90116-9}, \href
  {https://ui.adsabs.harvard.edu/abs/1988Icar...74..472K} {74, 472}

\bibitem[\protect\citeauthoryear{{Kasting}, {Pollack}  \& {Ackerman}}{{Kasting}
  et~al.}{1984}]{1984Icar...57..335K}
{Kasting} J.~F.,  {Pollack} J.~B.,   {Ackerman} T.~P.,  1984, \mn@doi [\icarus]
  {10.1016/0019-1035(84)90122-2}, \href
  {https://ui.adsabs.harvard.edu/abs/1984Icar...57..335K} {57, 335}

\bibitem[\protect\citeauthoryear{{Kasting}, {Whitmire}  \&
  {Reynolds}}{{Kasting} et~al.}{1993}]{1993Icar..101..108K}
{Kasting} J.~F.,  {Whitmire} D.~P.,   {Reynolds} R.~T.,  1993, \mn@doi
  [\icarus] {10.1006/icar.1993.1010}, \href
  {https://ui.adsabs.harvard.edu/abs/1993Icar..101..108K} {101, 108}

\bibitem[\protect\citeauthoryear{{Kasting}, {Chen}  \& {Kopparapu}}{{Kasting}
  et~al.}{2015}]{2015ApJ...813L...3K}
{Kasting} J.~F.,  {Chen} H.,   {Kopparapu} R.~K.,  2015, \mn@doi [\apjl]
  {10.1088/2041-8205/813/1/L3}, \href
  {https://ui.adsabs.harvard.edu/abs/2015ApJ...813L...3K} {813, L3}

\bibitem[\protect\citeauthoryear{{Kopparapu}, {Wolf}, {Arney}, {Batalha},
  {Haqq-Misra}, {Grimm}  \& {Heng}}{{Kopparapu}
  et~al.}{2017}]{2017ApJ...845....5K}
{Kopparapu} R.~k.,  {Wolf} E.~T.,  {Arney} G.,  {Batalha} N.~E.,  {Haqq-Misra}
  J.,  {Grimm} S.~L.,   {Heng} K.,  2017, \mn@doi [\apj]
  {10.3847/1538-4357/aa7cf9}, \href
  {https://ui.adsabs.harvard.edu/abs/2017ApJ...845....5K} {845, 5}

\bibitem[\protect\citeauthoryear{{Larson}, {Golaz}  \& {Cotton}}{{Larson}
  et~al.}{2002}]{2002JAtS...59.3519L}
{Larson} V.~E.,  {Golaz} J.-C.,   {Cotton} W.~R.,  2002, \mn@doi [Journal of
  Atmospheric Sciences] {10.1175/1520-0469(2002)059<3519:SSAMVI>2.0.CO;2},
  \href {https://ui.adsabs.harvard.edu/abs/2002JAtS...59.3519L} {59, 3519}

\bibitem[\protect\citeauthoryear{{Li}, {Wang}, {Guo}, {Yang}, {Xu}, {Han}  \&
  {Sun}}{{Li} et~al.}{2022}]{2022JAMES..1403127L}
{Li} T.,  {Wang} M.,  {Guo} Z.,  {Yang} B.,  {Xu} Y.,  {Han} X.,   {Sun} J.,
  2022, \mn@doi [Journal of Advances in Modeling Earth Systems]
  {10.1029/2022MS003127}, \href
  {https://ui.adsabs.harvard.edu/abs/2022JAMES..1403127L} {14, e2022MS003127}

\bibitem[\protect\citeauthoryear{{Linsenmeier}, {Pascale}  \&
  {Lucarini}}{{Linsenmeier} et~al.}{2015}]{2015P&SS..105...43L}
{Linsenmeier} M.,  {Pascale} S.,   {Lucarini} V.,  2015, \mn@doi [\planss]
  {10.1016/j.pss.2014.11.003}, \href
  {https://ui.adsabs.harvard.edu/abs/2015P&SS..105...43L} {105, 43}

\bibitem[\protect\citeauthoryear{{Liu} et~al.,}{{Liu}
  et~al.}{2012}]{2012GMD.....5..709L}
{Liu} X.,  et~al., 2012, \mn@doi [Geoscientific Model Development]
  {10.5194/gmd-5-709-2012}, \href
  {https://ui.adsabs.harvard.edu/abs/2012GMD.....5..709L} {5, 709}

\bibitem[\protect\citeauthoryear{{Liu}, {Ma}, {Wang}, {Tilmes}, {Singh},
  {Easter}, {Ghan}  \& {Rasch}}{{Liu} et~al.}{2016}]{2016GMD.....9..505L}
{Liu} X.,  {Ma} P.~L.,  {Wang} H.,  {Tilmes} S.,  {Singh} B.,  {Easter} R.~C.,
  {Ghan} S.~J.,   {Rasch} P.~J.,  2016, \mn@doi [Geoscientific Model
  Development] {10.5194/gmd-9-505-2016}, \href
  {https://ui.adsabs.harvard.edu/abs/2016GMD.....9..505L} {9, 505}

\bibitem[\protect\citeauthoryear{{Long} et~al.,}{{Long}
  et~al.}{2021}]{2021JAMES..1302647L}
{Long} M.~C.,  et~al., 2021, \mn@doi [Journal of Advances in Modeling Earth
  Systems] {10.1029/2021MS002647}, \href
  {https://ui.adsabs.harvard.edu/abs/2021JAMES..1302647L} {13, e02647}

\bibitem[\protect\citeauthoryear{{Lopez} et~al.,}{{Lopez}
  et~al.}{2019}]{2019A&A...631A..90L}
{Lopez} T.~A.,  et~al., 2019, \mn@doi [\aap] {10.1051/0004-6361/201936267},
  \href {https://ui.adsabs.harvard.edu/abs/2019A&A...631A..90L} {631, A90}

\bibitem[\protect\citeauthoryear{{Marsh}, {Mills}, {Kinnison}, {Lamarque},
  {Calvo}  \& {Polvani}}{{Marsh} et~al.}{2013}]{2013JCli...26.7372M}
{Marsh} D.~R.,  {Mills} M.~J.,  {Kinnison} D.~E.,  {Lamarque} J.-F.,  {Calvo}
  N.,   {Polvani} L.~M.,  2013, \mn@doi [Journal of Climate]
  {10.1175/JCLI-D-12-00558.1}, \href
  {https://ui.adsabs.harvard.edu/abs/2013JCli...26.7372M} {26, 7372}

\bibitem[\protect\citeauthoryear{{Mills} et~al.,}{{Mills}
  et~al.}{2016}]{2016JGRD..121.2332M}
{Mills} M.~J.,  et~al., 2016, \mn@doi [Journal of Geophysical Research
  (Atmospheres)] {10.1002/2015JD024290}, \href
  {https://ui.adsabs.harvard.edu/abs/2016JGRD..121.2332M} {121, 2332}

\bibitem[\protect\citeauthoryear{{Mlawer}, {Taubman}, {Brown}, {Iacono}  \&
  {Clough}}{{Mlawer} et~al.}{1997}]{1997JGR...10216663M}
{Mlawer} E.~J.,  {Taubman} S.~J.,  {Brown} P.~D.,  {Iacono} M.~J.,   {Clough}
  S.~A.,  1997, \mn@doi [\jgr] {10.1029/97JD00237}, \href
  {https://ui.adsabs.harvard.edu/abs/1997JGR...10216663M} {102, 16,663}

\bibitem[\protect\citeauthoryear{{Ohno} \& {Zhang}}{{Ohno} \&
  {Zhang}}{2019}]{2019ApJ...874....1O}
{Ohno} K.,  {Zhang} X.,  2019, \mn@doi [\apj] {10.3847/1538-4357/ab06cc}, \href
  {https://ui.adsabs.harvard.edu/abs/2019ApJ...874....1O} {874, 1}

\bibitem[\protect\citeauthoryear{{Oinas}, {Lacis}, {Rind}, {Shindell}  \&
  {Hansen}}{{Oinas} et~al.}{2001}]{2001GeoRL..28.2791O}
{Oinas} V.,  {Lacis} A.~A.,  {Rind} D.,  {Shindell} D.~T.,   {Hansen} J.~E.,
  2001, \mn@doi [\grl] {10.1029/2001GL013137}, \href
  {https://ui.adsabs.harvard.edu/abs/2001GeoRL..28.2791O} {28, 2791}

\bibitem[\protect\citeauthoryear{{Palubski}, {Shields}  \&
  {Deitrick}}{{Palubski} et~al.}{2020}]{2020ApJ...890...30P}
{Palubski} I.~Z.,  {Shields} A.~L.,   {Deitrick} R.,  2020, \mn@doi [\apj]
  {10.3847/1538-4357/ab66b2}, \href
  {https://ui.adsabs.harvard.edu/abs/2020ApJ...890...30P} {890, 30}

\bibitem[\protect\citeauthoryear{{Park}, {Bretherton}  \& {Rasch}}{{Park}
  et~al.}{2014}]{2014JCli...27.6821P}
{Park} S.,  {Bretherton} C.~S.,   {Rasch} P.~J.,  2014, \mn@doi [Journal of
  Climate] {10.1175/JCLI-D-14-00087.1}, \href
  {https://ui.adsabs.harvard.edu/abs/2014JCli...27.6821P} {27, 6821}

\bibitem[\protect\citeauthoryear{{Quirrenbach}}{{Quirrenbach}}{2022}]{2022RNAAS...6...56Q}
{Quirrenbach} A.,  2022, \mn@doi [Research Notes of the American Astronomical
  Society] {10.3847/2515-5172/ac5f0d}, \href
  {https://ui.adsabs.harvard.edu/abs/2022RNAAS...6...56Q} {6, 56}

\bibitem[\protect\citeauthoryear{{Richter} et~al.,}{{Richter}
  et~al.}{2022}]{2022WtFor..37..797R}
{Richter} J.~H.,  et~al., 2022, \mn@doi [Weather and Forecasting]
  {10.1175/WAF-D-21-0163.1}, \href
  {https://ui.adsabs.harvard.edu/abs/2022WtFor..37..797R} {37, 797}

\bibitem[\protect\citeauthoryear{{Rogers}}{{Rogers}}{2015}]{2015ApJ...801...41R}
{Rogers} L.~A.,  2015, \mn@doi [\apj] {10.1088/0004-637X/801/1/41}, \href
  {https://ui.adsabs.harvard.edu/abs/2015ApJ...801...41R} {801, 41}

\bibitem[\protect\citeauthoryear{Serdyuchenko, Gorshelev, Weber, Chehade  \&
  Burrows}{Serdyuchenko et~al.}{2014}]{amt-7-625-2014}
Serdyuchenko A.,  Gorshelev V.,  Weber M.,  Chehade W.,   Burrows J.~P.,  2014,
  \mn@doi [Atmospheric Measurement Techniques] {10.5194/amt-7-625-2014}, 7, 625

\bibitem[\protect\citeauthoryear{{Villanueva}, {Smith}, {Protopapa}, {Faggi}
  \& {Mandell}}{{Villanueva} et~al.}{2018}]{2018JQSRT.217...86V}
{Villanueva} G.~L.,  {Smith} M.~D.,  {Protopapa} S.,  {Faggi} S.,   {Mandell}
  A.~M.,  2018, \mn@doi [\jqsrt] {10.1016/j.jqsrt.2018.05.023}, \href
  {https://ui.adsabs.harvard.edu/abs/2018JQSRT.217...86V} {217, 86}

\bibitem[\protect\citeauthoryear{{Way} \& {Georgakarakos}}{{Way} \&
  {Georgakarakos}}{2017}]{2017ApJ...835L...1W}
{Way} M.~J.,  {Georgakarakos} N.,  2017, \mn@doi [\apjl]
  {10.3847/2041-8213/835/1/L1}, \href
  {https://ui.adsabs.harvard.edu/abs/2017ApJ...835L...1W} {835, L1}

\bibitem[\protect\citeauthoryear{{Williams} \& {Pollard}}{{Williams} \&
  {Pollard}}{2002}]{2002IJAsB...1...61W}
{Williams} D.~M.,  {Pollard} D.,  2002, \mn@doi [International Journal of
  Astrobiology] {10.1017/S1473550402001064}, \href
  {https://ui.adsabs.harvard.edu/abs/2002IJAsB...1...61W} {1, 61}

\bibitem[\protect\citeauthoryear{{Winters}, {Cloutier}, {Medina}, {Irwin},
  {Charbonneau}  \& {LTT 1445A planet mass paper co-authors}}{{Winters}
  et~al.}{2022}]{2022BAAS...54e.417W}
{Winters} J.,  {Cloutier} R.,  {Medina} A.,  {Irwin} J.,  {Charbonneau} D.,
  {LTT 1445A planet mass paper co-authors} 2022, in Bulletin of the American
  Astronomical Society. p. 102.417

\bibitem[\protect\citeauthoryear{{Wolf} \& {Toon}}{{Wolf} \&
  {Toon}}{2015}]{2015JGRD..120.5775W}
{Wolf} E.~T.,  {Toon} O.~B.,  2015, \mn@doi [Journal of Geophysical Research
  (Atmospheres)] {10.1002/2015JD023302}, \href
  {https://ui.adsabs.harvard.edu/abs/2015JGRD..120.5775W} {120, 5775}

\bibitem[\protect\citeauthoryear{{Wolf}, {Shields}, {Kopparapu}, {Haqq-Misra}
  \& {Toon}}{{Wolf} et~al.}{2017}]{2017ApJ...837..107W}
{Wolf} E.~T.,  {Shields} A.~L.,  {Kopparapu} R.~K.,  {Haqq-Misra} J.,   {Toon}
  O.~B.,  2017, \mn@doi [\apj] {10.3847/1538-4357/aa5ffc}, \href
  {https://ui.adsabs.harvard.edu/abs/2017ApJ...837..107W} {837, 107}

\bibitem[\protect\citeauthoryear{{Wordsworth} \& {Kreidberg}}{{Wordsworth} \&
  {Kreidberg}}{2021}]{2021arXiv211204663W}
{Wordsworth} R.,  {Kreidberg} L.,  2021, arXiv e-prints, \href
  {https://ui.adsabs.harvard.edu/abs/2021arXiv211204663W} {p. arXiv:2112.04663}

\bibitem[\protect\citeauthoryear{{Zeng} et~al.,}{{Zeng}
  et~al.}{2022}]{2022JGRD..12736452Z}
{Zeng} G.,  et~al., 2022, \mn@doi [Journal of Geophysical Research
  (Atmospheres)] {10.1029/2022JD036452}, \href
  {https://ui.adsabs.harvard.edu/abs/2022JGRD..12736452Z} {127, e2022JD036452}

\makeatother
\end{thebibliography}

\bsp	
\label{lastpage}
\end{document}